\documentclass[11pt]{article}
\usepackage{amsfonts}

\evensidemargin\oddsidemargin \topmargin2cm
\begin{document}

\begin{center}
\textbf{DECOHERENCE AND INFRARED DIVERGENCE}\footnote{%
Talk presented at the \textit{Second Winter Institute on
Foundations of Quantum Theory and Quantum Optics}. S N Bose
National Centre of Basic Sciences, Kolkata, India, January 2 - 11,
2002.}

\vspace{0.5cm}

Joachim Kupsch \\[0pt]

Fachbereich Physik, Universit\"at Kaiserslautern\\[0pt]
D-67653 Kaiserslautern, Germany
\end{center}

\begin{abstract}
The dynamics of a particle which is linearly coupled to the
environment of a Boson field is investigated. For velocity
coupling and for local (position) coupling the reduced dynamics of
the particle induces superselection rules for the momentum, if the
Boson field is infrared divergent. Thereby the Hamiltonian of the
total system remains bounded from below.
\end{abstract}

\section{Introduction}

Decoherence and superselection rules are the basis to understand how
classical physics can emerge within quantum theory. The mathematical
structure of quantum mechanics and of quantum field theory provides only a
few superselection rules, the most important being the charge superselection
rule related to the gauge invariance of the electromagnetic field, see e.g.
\cite{Wightman:1995} and the references therein. But there are definitively
not enough of these superselection rules to understand the classical
appearance of the world within quantum theory. A possible solution of this
problem is the emergence of superselection rules due to decoherence caused
by the interaction with the environment \cite{GJKKSZ:1996}.

We investigate the dynamics of a particle which is linearly
coupled to a Boson field. The Hamiltonian of the total system is
bounded from below. This talk concentrates on a specific aspect of
decoherence: decoherence due to infrared divergence of the Boson
field. In the case of velocity coupling the infrared divergence
induces a superselection rule for the momentum with uniform
estimates for the emergence of the superselection sectors \cite
{Kupsch:2000a}. For the model with local coupling the infrared
divergence also causes a strong decoherence between states of
different momenta, but the superselection sectors are not
uniformly induced. In this talk only some indications are given
about how these results can be derived. The detailed calculations
will be published in \cite{Kupsch/Smolyanov:2002}.

The models considered here have quadratic Hamiltonians, and they
can be solved in an explicit way. Despite of their unrealistic
simplicity such Hamiltonians have been successfully used to
construct heat bath models \cite {FKM:1965} and for understanding
the approach to thermal equilibrium \cite {Davies:1972}
\cite{Davies:1973}, or for calculating decoherence effects due to
radiation \cite{Duerr/Spohn:2000}. The structure of quadratic
Hamiltonians is therefore rich enough for obtaining interesting
results.

\section{The models\label{model}}

We consider a massive spinless particle coupled to a Boson field
of mass zero. The method presented here can be applied to a
particle in $\mathbb{R}^{d}$ with $d\geq 1$. But to simplify the
notations we restrict ourselves to one dimension. The Hilbert
space of the theory is the tensor space \newline
$\mathcal{H}_{S}\otimes
\mathcal{H}_{F}=\mathcal{L}^{2}(\mathbb{R})\otimes
\mathcal{F}(\mathcal{H}_{0})$ of the Hilbert space for the particle $%
\mathcal{H}_{S}=\mathcal{L}^{2}(\mathbb{R})$ and of the Fock space $\mathcal{H}%
_{F}=\mathcal{F}(\mathcal{H}_{0})$ for the Boson field, where $\mathcal{H}%
_{0}$ is the one-particle Hilbert space of the Bosons. The
Hamiltonian without interaction between the particle and the
field,
\begin{equation}
H_{(0)}=\frac{1}{2}\left( P^{2}+\omega ^{2}Q^{2}\right) \otimes
I_{F}+I_{S}\otimes H_{F}  \label{m2}
\end{equation}
is the sum of the Hamiltonian of a harmonic oscillator or free particle ($%
\omega =0$) on the Hilbert space $\mathcal{L}^{2}(\mathbb{R})$ and
the
Hamiltonian $H_{F}$ of a free field on the Fock space $\mathcal{F}(\mathcal{H%
}_{0})$. As interaction we take \newline
-- a velocity coupling $H_{P}=P\otimes \Phi (h)$ without harmonic force ($%
\omega =0$), \newline
-- a local interaction $H_{Q}=Q\otimes \Phi (h)$ together with a harmonic
force ($\omega >0$), \newline
where $\Phi (h)$ is the Boson field operator smeared with a suitably chosen
test function $h$.\newline
For these quadratic Hamiltonians the field equations are linear and can be
solved explicitly. To investigate the reduced dynamics we use the Heisenberg
picture and take the algebra generated by the Weyl operators as algebra of
observables.

\section{Weyl operators and classical phase space\label{Weyl}}

\subsection{Quantum mechanics of the particle}

As Hilbert space of the particle we take $\mathcal{H}_{S}=\mathcal{L}^{2}(%
\mathbb{R})$. The Weyl operators are defined as exponentials of
the canonical
variables, the position operator $Q$ and the momentum operator $P$, as $%
W_{S}(a,b)=\exp i\left( aP+bQ\right) $ with the parameters $(a,b)\in \mathbb{R}%
^{2}$. In the momentum representation on the Hilbert space $\mathcal{L}^{2}(%
\mathbb{R})$ the Weyl operators are given by
\begin{equation}
\left( W_{S}(a,b)f\right) (k)=\exp \left( \frac{i}{2}ab+iak\right) f(k-b).
\label{q3}
\end{equation}
From this identity follows that the arguments $(a,b)\in
\mathbb{R}^{2}$ can be interpreted as the position and the
momentum variables of the classical phase space of the particle.
The closure of the linear span of Weyl operators with the operator
norm is the Weyl algebra. This algebra is
smaller than the algebra of all bounded operators on $\mathcal{L}^{2}(\mathbb{R}%
)$. But the closure of the Weyl algebra with respect to the weak or the
strong operator topology includes all bounded operators.

For a one-dimensional harmonic oscillator with Hamiltonian $H=\frac{1}{2}%
\left( P^{2}+\omega ^{2}Q^{2}\right) ,\newline
\omega >0$, and unitary group $U(t)=\exp (-iHt)$ the time dependence of the
canonical variables is $U^{+}(t)PU(t)=P\cos \omega t-Q\omega \sin \omega t$
and $U^{+}(t)QU(t)=P\omega ^{-1}\sin \omega t+Q\cos \omega t$. The dynamics
of the Weyl operators then follows as $%
U^{+}(t)W_{S}(a,b)U(t)=W_{S}(a(t),b(t))$ with
\begin{equation}
\left(
\begin{array}{l}
a(t) \\
b(t)
\end{array}
\right) =R(t)\left(
\begin{array}{l}
a \\
b
\end{array}
\right) ,\;R(t)=\left(
\begin{array}{cc}
\cos \omega t & \omega ^{-1}\sin \omega t \\
-\omega \sin \omega t & \cos \omega t
\end{array}
\right)   \label{q5}
\end{equation}
where $R(t)$ is a group of rotations on $\mathbb{R}^{2}$. For
$\omega \rightarrow +0$ the transformations $R(t)$ have a well
defined limit, which corresponds to the dynamics $P(t)=P(0)=P$ and
$Q(t)=Q(0)+Pt$ or $a(t)=a+bt$ and $b(t)=b$ of a free particle.

\subsection{Field operators\label{fock}}

The basic Hilbert space for the Boson field is $\mathcal{H}_{0}=\mathcal{L}%
^{2}(\mathbb{R}^{n})$ with inner product $\left\langle f\mid
g\right\rangle
=\int \overline{f(k)}g(k)d^{n}k$ and norm $\left\| f\right\| =\sqrt{%
\left\langle f\mid f\right\rangle }$. The operator $M$ is the one-particle
Hamilton operator $\left( Mf\right) (k)=\omega (k)~f(k)$ with $\omega (k)=%
\sqrt{k^{2}},$~$k^{2}=\sum_{\mu =1}^{n}k_{\mu
}^{2},\;k=(k_{1},...,k_{n})\in \mathbb{R}^{n}.$ The operator $M$
and its inverse $M^{-1}$ are unbounded
operators. To define real vectors we introduce the involution $f^{\ast }(k):=%
\overline{f(-k)}.$ The operator $M$ is a real operator $(Mf)^{\ast
}=Mf^{\ast }$ with respect to this involution.

The one-particle space $\mathcal{H}_{0}$ generates the Fock space $\mathcal{F%
}(\mathcal{H}_{0})$. Let us denote with $f\circ g$ the symmetric tensor
product. Then the normalizations of the Fock space $\mathcal{F}(\mathcal{H}%
_{0})$ are chosen such that the inner product of exponential vectors $\exp
f=1_{vac}+f+\frac{1}{2}f\circ f+...\in \mathcal{F}(\mathcal{H}_{0}),~f\in
\mathcal{H}_{0}$, is $\left\langle \exp f\mid \exp g\right\rangle =\exp
\left\langle f\mid g\right\rangle $. The normalized exponential vector $\exp
\left( f-\frac{1}{2}\left\| f\right\| ^{2}\right) ,~\,f\in \mathcal{H}_{0}$,
is usually called a coherent state. We use the notation $\sigma (f)$ for the
projection operator onto that state. \newline
The creation operators: $a^{+}(h),~h\in \mathcal{H}_{0}$, are uniquely
defined by $a^{+}(h)\exp f:=h\circ \exp f$, and the annihilation operators $%
a(h),~h\in \mathcal{H}_{0}$ are the adjoint operators $a(h)=\left(
a^{+}(h)\right) ^{+}$. These operators are normalized to $\left[
a(f),~a^{+}(g)\right] =\left\langle f\mid g\right\rangle $. \newline
The Hamilton operator $H_{F}$ of the free field is uniquely determined by $%
H_{F}\exp f=Mf\circ \exp f$. The Hilbert space of real vectors
$f=f^{\ast }\in \mathcal{H}_{0},~$is denoted by
$\mathcal{H}_{\mathbb{R}}$. It is convenient to introduce a test
function space $\mathcal{E}_{\mathbb{R}}\subset
\mathcal{H}_{\mathbb{R}}$ on which arbitrary powers of $M$ are
defined, \newline $\mathcal{E}_{\mathbb{R}}=\left\{ f\in
\mathcal{H}_{\mathbb{R}}\mid M^{n}f\in
\mathcal{H}_{\mathbb{R}},~n=-1,0,1,...\right\} $. For real test
functions $f\in \mathcal{E}_{\mathbb{R}}$ we define the\textit{\
field operator }$\Phi (f)$ and
the \textit{conjugate momentum field }$\Pi (f)$%
\begin{equation}
\begin{array}{c}
\Phi (f):=\frac{1}{\sqrt{2}}\left( a^{+}(M^{-\frac{1}{2}}f)+a(M^{-\frac{1}{2}%
}f)\right)  \\
\Pi (f):=i\left[ H,\Phi (f)\right] =\frac{i}{\sqrt{2}}\left( a^{+}(M^{\frac{1%
}{2}}f)-a(M^{\frac{1}{2}}f)\right)
\end{array}
\label{f2}
\end{equation}
see e. g. \cite{GJ:1987}. These fields are self-adjoint and they satisfy the
canonical commutations relations $\left[ \Phi (f),\Phi (g)\right] =\left[
\Pi (f),\Pi (g)\right] =0\;\mathrm{and}\;\left[ \Phi (f),\Pi (g)\right]
=i\,\left\langle f\mid g\right\rangle .$ From these definitions follows that
the arguments of the canonical fields $\Pi (u)$ and $\Phi (v)$ can be
extended to vectors $u\in \mathcal{H}_{-1\mathbb{R}}$ and $v\in \mathcal{H}_{1%
\mathbb{R}}$, where $\mathcal{H}_{\pm 1\mathbb{R}}$ are the real
Hilbert spaces with the inner products $\left\langle f\mid
g\right\rangle _{\pm
1}=\left\langle f\mid M^{\mp 1}g\right\rangle $. The spaces $\mathcal{H}_{1%
\mathbb{R}}$ and $\mathcal{H}_{-1\mathbb{R}}$ are dual spaces with
respect to the
inner product $\left\langle f\mid g\right\rangle $ of $\mathcal{H}_{\mathbb{R}}$%
.

The Hamiltonians $M$ and $H_{F}$ generate the unitary groups $U_{0}(t)=\exp
(-iM\,t)$ on $\mathcal{H}_{0}$ and $U_{F}(t)=\exp (-iH_{F}t)$ on $\mathcal{F}%
(\mathcal{H}_{0})$. The time evolution of the field $\Phi (v)$ and of the
canonical momentum field $\Pi (u)$ is
\begin{equation}
\begin{array}{l}
\Phi (v,t)=U_{F}^{+}(t)\Phi (v)U_{F}(t)=\Phi \left( \cos (M\,t)v\right) +\Pi
\left( M^{-1}\sin (M\,t)v\right)  \\
\Pi (u,t)=U_{F}^{+}(t)\Pi (u)U_{F}(t)=-\Phi \left( M\sin (M\,t)u\right) +\Pi
(\cos (M\,t)u).
\end{array}
\label{f7}
\end{equation}
These identities are defined for test functions $(u,v)\in \mathcal{H}_{-1%
\mathbb{R}}\times \mathcal{H}_{1\mathbb{R}}$.

\subsection{Weyl operators in QFT\label{Weylfield}}

We define the unitary Weyl operators $W_{F}(u,v),~(u,v)\in \mathcal{H}_{-1%
\mathbb{R}}\times \mathcal{H}_{1\mathbb{R}}$, by
\begin{equation}
W_{F}(u,v):=\exp \left( i\Pi (u)+i\Phi (v)\right) .  \label{f8}
\end{equation}
The $\mathbb{R}$-linear space of the arguments $(u,v)\in \mathcal{H}_{-1\mathbb{R}%
}\times \mathcal{H}_{1\mathbb{R}}$ can be interpreted as the
classical phase space of the theory. The expectation of the Weyl
operators between exponential vectors is
\begin{eqnarray}
&&\left\langle \exp f\mid W_{F}(u,v)\exp f\right\rangle   \nonumber \\
&=&\exp \left( \left\| f\right\| ^{2}-\frac{1}{4}\left\| u\right\| _{-1}^{2}-%
\frac{1}{4}\left\| v\right\| _{1}^{2}+i\mathrm{Im}\left\langle M^{\frac{1}{2}%
}u+iM^{-\frac{1}{2}}v\mid f\right\rangle \right)   \label{f13}
\end{eqnarray}
which includes the vacuum expectation. The expectation of $W_{F}(u,v)$ in a
K(ubo) M(artin) S(chwinger) state of inverse temperature $\beta >0$ is
\footnote{%
The KMS states are temperature states on the Weyl algebra. They are defined
also for Hamiltonians with a continuous spectrum, see \cite{Araki/Woods:1963}%
.}
\begin{equation}
\left\langle W_{F}(u,v)\right\rangle _{\beta }=\exp \left( -\frac{1}{4}%
\left\langle u\mid \coth \frac{\beta M}{2}u\right\rangle _{-1}-\frac{1}{4}%
\left\langle v\mid \coth \frac{\beta M}{2}v\right\rangle _{1}\right) .
\label{f15}
\end{equation}

The dynamics of the Weyl operators follows immediately from (\ref{f7}) as
\newline
$U_{F}^{+}(t)W_{F}(u,v)U_{F}(t)=W_{F}\left( u(t),v(t)\right) $ with
\begin{equation}
\left(
\begin{array}{c}
u(t) \\
v(t)
\end{array}
\right) =R(t)\left(
\begin{array}{c}
u \\
v
\end{array}
\right) ,\;R(t)=\left(
\begin{array}{cc}
\cos (M\,t) & M^{-1}\sin (M\,t) \\
-M\sin (M\,t) & \cos (M\,t)
\end{array}
\right)   \label{f16}
\end{equation}
The mapping $R(t)$ is a rotation of the phase space $\mathcal{H}_{-1\mathbb{R}%
}\times \mathcal{H}_{1\mathbb{R}}$ with respect to the quadratic form $%
\left\langle u\mid M^{\frac{1}{2}}u\right\rangle +\left\langle v\mid M^{-%
\frac{1}{2}}v\right\rangle =\left\| u\right\| _{-1}^{2}+\left\| v\right\|
_{1}^{2}$ of the vacuum functional.

\section{Dynamics of the system particle and field}

\subsection{The flow on the classical phase space\label{particlefield}}

The Hilbert space of the total system particle and field is $\mathcal{H}%
_{S}\otimes \mathcal{H}_{F}=\mathcal{L}^{2}(\mathbb{R})\otimes \mathcal{F}(%
\mathcal{H}_{0})$. The classical phase space can be represented by $%
(a,u,b,v)\in \mathbb{R}\times \mathcal{H}_{-1\mathbb{R}}\times
\mathbb{R}\times \mathcal{H}_{1\mathbb{R}}$ and the Weyl operators
are
\begin{equation}
\begin{array}{c}
W(a,u,b,v)=\exp i\left( aP+bQ\right) \otimes I_{F}+I_{S}\otimes \left( \Pi
(u)+\Phi (v)\right)  \\
=W_{S}(a,b)\otimes W_{F}(u,v)\;\mathrm{with}\;(a,u,b,v)\in \mathbb{R}\mathbf{%
\times }\mathcal{H}_{-1\mathbb{R}}\times \mathbb{R}\mathbf{\times }\mathcal{H}_{1%
\mathbb{R}}.
\end{array}
\label{p1}
\end{equation}
The dynamics of the fields is uniquely determined by the dynamics of the
Weyl operators
\begin{equation}
U^{+}(t)W(a,u,b,v)U(t)=W(a(t),u(t),b(t),v(t)).  \label{p2}
\end{equation}
where the flow $(a(t),u(t),b(t),v(t))$ on the classical phase space $\mathbb{R}%
\times \mathcal{H}_{-1\mathbb{R}}\times \mathbb{R}\times
\mathcal{H}_{1\mathbb{R}}$ is given by a one-parameter group
$R(t)=\exp Lt$ of symplectic transformations. The generator $L$ of
the group $R(t)$ is defined on the
restricted phase space $\mathbb{R}\times \mathcal{E}_{\mathbb{R}}\times \mathbb{R}%
\times \mathcal{E}_{\mathbb{R}}$ and can be read off from the
linear field equations of the theory.

\subsection{Reduced dynamics}

We assume that the initial state of the total system is given by
\begin{equation}
\rho =\rho _{S}\otimes \rho _{F}\in \mathcal{D}(\mathcal{H}_{S}\otimes
\mathcal{H}_{F})  \label{r1}
\end{equation}
where $\rho _{S}\in \mathcal{D}(\mathcal{H}_{S})$ is the initial state of
the particle and $\rho _{F}\in \mathcal{D}(\mathcal{H}_{F})$ is the
reference state of the field. The dynamics of any observable $A\in \mathcal{B%
}(\mathcal{H}_{S})$ of the particle is then
\begin{equation}
A\rightarrow A(t)=\Phi _{t}\left[ A\right] :=\mathrm{tr}_{F}U^{+}(t)\left(
A\otimes I_{F}\right) U(t)\rho _{F}.  \label{r2}
\end{equation}

If $A$ is the Weyl operator $W_{S}(a,b)$ of the particle, then $%
W_{S}(a,b)\otimes I_{F}=W(a,0,b,0)$ is a Weyl operator of the
total system as defined in Sect. 4.1 with \newline
$(a,u=0,b,v=0)\in \mathbb{R}\times
\mathcal{H}_{-1\mathbb{R}}\times \mathbb{R}\times
\mathcal{H}_{1\mathbb{R}}$. For quadratic Hamiltonians the time evolution $%
U^{+}(t)\left( W_{S}(a,b)\otimes I_{F}\right) U(t)$ yields the Weyl operator
(\ref{p2}) $W(a(t),u(t),b(t),v(t))$. Thereby the trajectories $%
(a(t),u(t),b(t),v(t))$ start from the initial conditions \newline
$(a(0),u(0),b(0),v(0))=(a,0,b,0)$. Since $W(a,u,b,v)=W_{S}(a,b)\otimes
W_{F}(u,v)$ the reduced dynamics (\ref{r2}) is simply calculated as
\begin{equation}
\Phi _{t}\left[ W_{S}(a,b)\right] =W_{S}(a(t),b(t))~\chi (a,b;t)  \label{r3}
\end{equation}
with the function $\chi (a,b;t)=\mathrm{tr}_{F}W_{F}(u(t),v(t))\rho _{F}$. A
decrease of the trace $\chi (a,b;t)$ for large $t$ indicates decoherence
effects due to the environment of the Bosons.

If $\rho _{F}$ is the projection operator $\sigma (g)$ onto the coherent
state $\exp \left( g-\frac{1}{2}\left\| g\right\| ^{2}\right) ,g\in \mathcal{%
H}_{0}$, the trace $\chi (a,b;t)=\mathrm{tr}_{F}W_{F}(u(t),v(t))\sigma (g)$
is bounded by, see (\ref{f13}),
\begin{equation}
\left| \chi (a,b;t)\right| =\exp \left( -\frac{1}{4}\left\| u(t)\right\|
_{-1}^{2}-\frac{1}{4}\left\| v(t)\right\| _{1}^{2}\right) .  \label{r5}
\end{equation}
The choice of the coherent states simplifies the calculations considerably.
But the statements about decoherence are qualitatively not affected, if one
takes another statistical operator as state of the environment \cite
{Kupsch:2000}.

If the Boson field is in a state of inverse temperature $\beta >0$ the
function $\chi (a,b;t)$ is given by (\ref{f15})
\begin{equation}
\begin{array}{c}
\chi _{\beta }(t)=\exp \left( -\frac{1}{4}\left\langle u(t)\mid \coth \frac{%
\beta M}{2}u(t)\right\rangle _{-1}-\frac{1}{4}\left\langle v(t)\mid \coth
\frac{\beta M}{2}v(t)\right\rangle _{1}\right) \\
<\exp \left( -\frac{1}{4}\left\| u(t)\right\| _{-1}^{2}-\frac{1}{4}\left\|
v(t)\right\| _{1}^{2}\right) .
\end{array}
\label{r6}
\end{equation}

\textbf{Remark }The technique to calculate the dynamics of a subsystem with
the help of Weyl operators has been used by Davies to derive the approach of
thermal equilibrium in a heat bath \cite{Davies:1972}.

\subsection{Evaluation of the dynamics and induced superselection rules}

\subsubsection{Independent systems}

We first consider a free or harmonically bound particle on the real line and
a free field with Hamiltonian (\ref{m2}). The operators
\begin{equation}
R_{0}(t)=\left(
\begin{array}{cc}
\cos (\widehat{M}_{0}\,t) & \widehat{M}_{0}^{-1}\sin (\widehat{M}_{0}\,t) \\
-\widehat{M}_{0}\sin (\widehat{M}_{0}\,t) & \cos (\widehat{M}_{0}\,t)
\end{array}
\right) \;\mathrm{with}\;\widehat{M}_{0}=\left(
\begin{array}{cc}
\omega  & 0 \\
0 & M
\end{array}
\right)   \label{p3}
\end{equation}
are rotations on the phase space $\mathbb{R}\times \mathcal{H}_{-1\mathbb{R}%
}\times \mathbb{R}\times \mathcal{H}_{1\mathbb{R}}$. This rotation
is the tensor product of (\ref{q5}) and (\ref{f16}).

\subsubsection{Interacting systems\label{interaction}}

\paragraph{Velocity coupling}

The Hamiltonian with the velocity coupling
\begin{equation}
\begin{array}{c}
H=\frac{1}{2}P^{2}\otimes I_{F}+P\otimes \Phi (h)+I_{S}\otimes H_{F} \\
=\frac{1}{2}\left( P\otimes I_{F}+I_{S}\otimes \Phi (h)\right)
^{2}+I_{S}\otimes \left( H_{F}-\frac{1}{2}\Phi ^{2}(h)\right) ^{2}
\end{array}
\label{v1}
\end{equation}
is bounded from below if $\left\| M^{-1}h\right\| \leq 1$. The momentum $P$
is a conserved quantity and it is possible to calculate the reduced dynamics
of this model \cite{Kupsch:2000a}. The Bosonic part of (\ref{v1}) is the van
Hove model \cite{Hove:1952}. If $h$ has low energy contributions such that $h
$ is no longer in the domain of the operator $M^{-\frac{3}{2}},~h\notin
\mathcal{D}(M^{-\frac{3}{2}})$, the ground state of the van Hove model
disappears in the continuum, see \cite{Schroer:1963} \cite
{Arai/Hirokawa:1999} Sect.6.1. Moreover, in the van Hove model and in our
model the mean Boson number diverges due to the soft Bosons. As a
consequence of this infrared catastrophe the selection rule of the momentum
of the particle becomes a superselection rule \cite{Kupsch:2000a}. Here we
want to understand this effect with the help of the flow on the classical
phase space. Assuming the stronger constraints
\begin{equation}
\left\| M^{-1}h\right\| <1,\;\mathrm{and}\;h\in \mathcal{D}(M^{-\frac{3}{2}%
})\subset \mathcal{H}_{\mathbb{R}},  \label{v4}
\end{equation}
the dynamics is unitarily equivalent to the free dynamics, and the flow can
be easily calculated, see \cite{Kupsch/Smolyanov:2002}. Starting from the
initial values $(a(0),u(0),b(0),v(0))=(a,0,b,0)$ we obtain
\begin{equation}
\left(
\begin{array}{c}
a(t) \\
u(t) \\
b(t) \\
v(t)
\end{array}
\right) =\left(
\begin{array}{c}
a+b\left\langle M^{-1}h\mid (\sin Mt)M^{-2}h\right\rangle +\alpha ^{2}bt \\
b(I-\cos Mt)M^{-2}h \\
b \\
b(\sin Mt)M^{-1}h
\end{array}
\right) .  \label{v5}
\end{equation}
Since $M$ has an absolutely continuous spectrum, the vector functions $%
u(t)\in \mathcal{H}_{-1\mathbb{R}}$ and $v(t)\in
\mathcal{H}_{1\mathbb{R}}$ remain bounded for ~$t\rightarrow
\infty $.

These expressions have been derived under the conditions (\ref{v4}). But the
identities (\ref{v5}) remain meaningful under the weaker condition $\left\|
M^{-1}h\right\| \leq 1$. If the low energy contributions in $h$ are strong
enough such that $h\notin \mathcal{D}(M^{-\frac{3}{2}})$, there is a
qualitative change in the time evolution: the Hamiltonian (\ref{v1}) is no
longer unitarily equivalent to the free Hamiltonian, and the norms
\begin{equation}
\begin{array}{c}
\left\| u(t)\right\| _{-1}=b^{2}\left\| (I-\cos Mt)M^{-2}h\right\|
_{-1}=b^{2}\left\| (I-\cos Mt)M^{-\frac{3}{2}}h\right\| \\
\left\| v(t)\right\| _{1}=b^{2}\left\| (\sin Mt)M^{-1}h\right\|
_{1}=b^{2}\left\| (\sin Mt)M^{-\frac{3}{2}}h\right\|
\end{array}
\label{v6}
\end{equation}
diverge for ~$t\rightarrow \infty $ if $b\neq 0$. As a consequence the
decoherence function decreases to zero in both the cases: an environment
given by coherent states (\ref{r5}) and an environment given by a thermal
state (\ref{r6}). The formulas (\ref{v5}) then imply the uniform bound
\begin{equation}
\left\| \Phi _{t}\left[ W_{S}(a,b)\right] \right\| \leq \exp \left(
-b^{2}\varphi (t)\right)  \label{v7}
\end{equation}
with a positive non-decreasing function $\varphi (t)$ which diverges for $%
t\rightarrow \infty $. If $b=0$ the relations (\ref{r3}) and (\ref{v5})
imply $\Phi _{t}\left[ W_{S}(a,b=0)\right] =W_{S}(a,b=0)$.

Let $\mathcal{L}^{2}(\mathbb{R})$ be the Hilbert space of the
particle in the momentum representation. For any interval
$\mathbf{I}\subset \mathbb{R}$ of the real line we denote with
$\mathrm{P}(\mathbf{I})$ the projection operator, which maps
$\mathcal{L}^{2}(\mathbb{R})$ into $\mathcal{L}^{2}(\mathbf{I})$,
i.e. $f(x)\in \mathcal{L}^{2}(\mathbb{R})$ is mapped onto $\left( \mathrm{P}(%
\mathbf{I})f\right) (x)=f(x)$ if $x\in \mathbf{I}$, and $\left( \mathrm{P}(%
\mathbf{I})f\right) (x)=0$ if $x\notin \mathbf{I}$. From (\ref{q3}) follows
\begin{equation}
\mathrm{P}(\mathbf{I}_{2})W_{S}(a,b)\mathrm{P}(\mathbf{I}_{1})=0\quad
\mathrm{if}\;b\notin \mathbf{I}_{2}-\mathbf{I}_{1}  \label{v8}
\end{equation}
for any two non overlapping intervals $\mathbf{I}_{1}$ and $\mathbf{I}_{2}$
of the real line. The relations (\ref{r3}) (\ref{v7}) and (\ref{v8}) yield $%
\left\| \mathrm{P}(\mathbf{I}_{2})\Phi _{t}\left[ W_{S}(a,b)\right] \mathrm{P%
}(\mathbf{I}_{1})\right\| \leq \exp \left( -\delta ^{2}\varphi (t)\right) $,
where $\delta =\mathrm{dist}(\mathbf{I}_{1},\mathbf{I}_{2})>0$ is the
distance between the intervals. Moreover any linear combination $A=\sum
c_{j}W_{S}(a_{j},b_{j})$ of Weyl operators satisfies an estimate $\left\|
\mathrm{P}(\mathbf{I}_{2})\Phi _{t}\left[ A\right] \mathrm{P}(\mathbf{I}%
_{1})\right\| \leq C_{A}\exp \left( -\delta ^{2}\varphi (t)\right) $ with
some constant $C_{A}$. These estimates imply: for all operators $A$ of the
Weyl algebra we have
\begin{equation}
\lim_{t\rightarrow \infty }\left\| \mathrm{P}(\mathbf{I}_{2})\Phi _{t}\left[
A\right] \mathrm{P}(\mathbf{I}_{1})\right\| =0,  \label{v10}
\end{equation}
and for all bounded observables $A$ the strong convergence
\begin{equation}
\lim_{t\rightarrow \infty }\mathrm{P}(\mathbf{I}_{2})\Phi _{t}\left[ A\right]
\mathrm{P}(\mathbf{I}_{1})f=0\;\mathrm{for~all}\;f\in \mathcal{L}^{2}(\mathbb{R}%
)  \label{v11}
\end{equation}
follows. Hence the reduced dynamics leads to a superselection rule for the
momentum.

\textbf{Remark }A uniform estimate of the type (\ref{v10}) has been derived
for all bounded observables $A$ with other methods in \cite{Kupsch:2000a}.

\paragraph{Position coupling}

For the Hamiltonian with the position coupling
\begin{equation}
H=\frac{1}{2}\left( P^{2}+\omega ^{2}Q^{2}\right) \otimes I_{F}+Q\otimes
\Phi (h)+I_{S}\otimes H_{F}  \label{p5}
\end{equation}
the flow on the phase space is -- as in the free case -- a rotation (\ref{p3}%
) $R(t)$. But the square of the energy operator $\widehat{M}$ is now given
by
\begin{equation}
\widehat{M}^{2}=\left(
\begin{array}{cc}
\omega ^{2} & \left\langle h\mid \cdot \right\rangle  \\
h & M^{2}
\end{array}
\right) .  \label{p6}
\end{equation}
The Hamiltonian (\ref{p5}) has a lower bound if and only if the operator (%
\ref{p6}) is positive. This condition is fulfilled if $h\in \mathcal{H}_{%
\mathbb{R}}$ satisfies the norm bound $\left\| M^{-1}h\right\|
\leq \omega .$ The operator (\ref{p6}) corresponds to the
Hamiltonian of the Friedrichs model, see \cite{Friedrichs:1948},
\cite{Exner:1985}, and Sect. 6.2 in \cite {Arai/Hirokawa:1999}. If
the function $h$ satisfies some smoothness and support
restrictions, the operator $\widehat{M}$ has an absolutely
continuous spectrum. Then one expects that either the harmonic
oscillator decays into the ground state (if the Boson field is in
a normal state), or the harmonic oscillator thermalizes into the
canonical ensemble (if the
Boson field is in a KMS state with positive temperature). But in the case $%
\left\| M^{-1}h\right\| ^{2}=\left\langle h\mid M^{-2}h\right\rangle =\omega
^{2}$ the spectrum of the operator $\widehat{M}$ includes zero. If, moreover
the low energy contributions dominate, the mean Boson number diverges and
strong decoherence effects emerge, which cause a non-uniform superselection
rule for the momentum of the particle. The details of the calculations will
be given in \cite{Kupsch/Smolyanov:2002}.

\end{document}